\documentclass[journal]{IEEEtran}

\usepackage{cite}
\usepackage{amsmath,amssymb,amsfonts}
\usepackage{algorithmic}
\usepackage{graphicx}
\usepackage{textcomp}

\usepackage{siunitx}
\usepackage{pgfplots}
\pgfplotsset{compat=1.14}
\usepackage{adjustbox}

\DeclareSIUnit{\pixel}{px}
\renewcommand{\vec}{\boldsymbol}

\makeatother

\begin{document}

\title{X-ray Scatter Estimation Using Deep Splines}
%
%
%

\author{
    Philipp~Roser, Annette~Birkhold, Alexander~Preuhs, Christopher~Syben, Lina~Felsner, Elisabeth~Hoppe, Norbert~Strobel, Markus~Korwarschik, Rebecca~Fahrig, Andreas~Maier
    \thanks{P.~Roser, A.~Preuhs, C.~Syben, L.~Felsner, E.~Hoppe, and A.~Maier are with the Pattern Recognition Lab, Department of Computer Science, Friedrich-Alexander Universit\"at Erlangen-N\"urnberg, Erlangen, Germany. P.~Roser is funded by the Erlangen Graduate School in Advanced Optical Technologies (SAOT), Friedrich-Alexander Universit\"at Erlangen-N\"urnberg, Erlangen, Germany. A.~Maier is principal investigator at the SAOT. A.~Birkhold, M.~Kowarschik, and R.~Fahrig are employees of Siemens Healthcare GmbH, 91301 Forchheim, Germany. N.~Strobel is with the Institute of Medical Engineering Schweinfurt, University of Applied Sciences W\"urzburg-Schweinfurt, 97421 Schweinfurt, Germany. }%
}%

\maketitle

\begin{abstract}
Algorithmic X-ray scatter compensation is a desirable technique in flat-panel X-ray imaging and cone-beam computed tomography.
State-of-the-art U-net based image translation approaches yielded promising results.
As there are no physics constraints applied to the output of the U-Net, it cannot be ruled out that it yields spurious results.
Unfortunately, those may be misleading in the context of medical imaging. 
To overcome this problem, we propose to embed B-splines as a known operator into neural networks.
This inherently limits their predictions to well-behaved and smooth functions.
In a study using synthetic head and thorax data as well as real
thorax phantom data, we found that our approach performed on par with U-net when comparing both algorithms based on quantitative performance metrics. However, our approach not only reduces runtime and parameter complexity, but we also found it much more robust to unseen noise levels.
While the U-net responded with visible artifacts, our approach preserved the X-ray signal's frequency characteristics.
\end{abstract}


%

\section{Introduction}
%
%
%
%
\IEEEPARstart{X}{ray} transmission imaging enables advanced diagnostic imaging or facilitates instrument guidance during minimally-invasive interventions.
Unfortunately, primary photons scattered by the patient's body impair image quality at the detector. 
Especially cone-beam X-ray systems using flat-panel detectors suffer from this phenomenon due to the large X-ray field and the resulting high scatter-to-primary ratio.
Besides contrast degradation in X-ray projections, artifacts in cone-beam computed tomography (CBCT) deteriorate the image quality in the reconstructed volume.
Since the advent of CBCT, various methods to compensate for scatter have been developed \cite{Ruehrnschopf:2011:ScatterCompensation1,Ruehrnschopf:2011:ScatterCompensation2}.
These can be categorized into physical and algorithmic scatter compensation methods. 

\subsection{Physical Scatter Compensation}
Physical scatter compensation refers to the direct manipulation of the X-ray field to either suppress scattered photons from reaching the detector, or modulate them to enable differentiation into scattered and primary photons.
A simple yet effective approach is to increase the patient to detector distance, the so-called air gap \cite{Groedel:1926:AirGap,Janus:1926:AirGap,Neitzel:1992:AirGap,Persliden:1997:AirGap}.
Since the number of scattered photons depends on the irradiated volume, the scatter-to-primary ratio can be reduced by shaping the X-ray field to a small volume of interest.
In slit scanning, the stitching of these smaller volumes of interest yields a normal-sized reconstruction \cite{Barnes:1991:Scatter,Bhagtani:2009:Slit}.
Although slit scanning reduces patient dose and scattered radiation, the prolonged acquisition time complicates motion compensation and X-ray tube heat management \cite{Schmidt:2004:SlitFeasibility}.
Since relying on large air gaps or strict collimation limits the flexibility of the imaging system, clinical systems are equipped with anti-scatter grids that physically block scattered photons \cite{Neitzel:1992:AirGap,Bucky:1913:ASG,Kalender:1982:ASG,Chan:1982:ASG,Chan:1985:ASG,Chan:1990:ASG}.
Although state-of-the-art for clinical CBCT, anti-scatter grids have disadvantages. 
For example, they can increase radiation exposure \cite{Aichinger:2004:IQ} and lead to Moir\'{e} and grid line artifacts \cite{Gauntt:2006:ASGLines}.
Overall, the efficiency and effectiveness of anti-scatter grids depends on multiple factors, e.g., air gap, photon energy, source-to-detector distance, and detector resolution \cite{Ruehrnschopf:2011:ScatterCompensation2,Singh:2014:ASGLines,Rana:2016:ASGLines}.
In contrast to detector-side grids, primary modulation follows a different operation principle \cite{BaniHashemi:2005:PrimaryModulation,Zhu:2006:PrimaryModulation,Bier:2017:PrimaryModulation}.
In the shadow of this modulator grid, almost no primary radiation is measured.
Instead, information of the scatter properties of the irradiated volume is encoded.
Using demodulation algorithms, the X-ray projection and the associated scatter distribution can be distinguished.
Due to its complex structure, primary modulation, especially with C-arm systems, has not made its way into the clinical practice.
In conclusion, hardware-based solutions induce additional manufacturing costs and limit flexibility.
Also, since most approaches need algorithmic support or look-up tables, software-only solutions are highly desirable \cite{Ruehrnschopf:2011:ScatterCompensation1,Ruehrnschopf:2011:ScatterCompensation2}.

\subsection{Algorithmic Scatter Compensation}
Algorithmic scatter compensation approaches typically try to estimate the scatter signal from the acquired X-ray projections first. 
The estimated scatter image is then subtracted to obtain the primary signal.
For CBCT, using Monte Carlo (MC) methods or Boltzmann transport solvers are common approaches \cite{Zbijewski:2004:MCScatter,Poludniowski:2009:MCScatter,Wang:2018:Acuros}.
Both require an initial reconstruction.
Their computational complexity can partially be coped with using dedicated hardware and variance reduction techniques \cite{Badal:09:MCGPU} or by using a coarse simulation to derive the scattering model online \cite{Baer:2012:HSE}.
Trading off effectiveness for efficiency, model-based approaches are typically preferred for clinical applications.
Such approaches rely on either simplified physical or analytical models \cite{Swindell:1996:PhysicalModel,Yao:2009:AnalyticalModel,Meyer:2009:EllipticModel} or convolution kernels \cite{Ohnesorge:1999:KSE,Li:2008:KSE,Sun:2010:KSE} in combination with an iterative correction scheme.
Although being fast, their ability to generalize to different imaging settings is limited.

Recently, learning-based methods found their way to modeling physical processes \cite{Chen:2018:PDEs,Roser:19:DoseLearning,roser2020simultaneous}. 
Especially the application of the U-net \cite{Ronneberger:2015:Unet} to the task of scatter estimation, referred to as deep scatter estimation, yields superior results compared to kernel-based or MC methods, respectively \cite{Maier:2019:DSE}.
The U-net based method has the potential to become the new gold standard in the domain of scatter correction.
However, there are several drawbacks to employing a deep U-net.
First, deep convolutional neural networks are challenging to comprehend in their operating principle.
Given the high parameter complexity of such a U-net, large amounts of training samples are mandatory to arrive at a robust solution.
Second, to maintain a fast performance, the U-net requires a dedicated top tier graphics processing unit (GPU), which might not be universally available.
Third, the U-net is a universal function approximator without including any known physical characteristics of scattered radiation.

To come up with an alternative, it appears attractive to build on the rich prior knowledge about X-ray scatter.
Also, it has been shown that the incorporation of prior knowledge into neural networks reduces error bounds \cite{Maier:2019:KnownOperators} and simplifies the analysis of the deployed network \cite{stimpel2019multi}.
Although subject to photon Poisson noise, X-ray scatter is mainly a low-frequency signal in the diagnostic energy regime \cite{Glover:1982:Compton,Boone:1988:Scatter}.
In a previous proof-of-concept study, we exploited this property by approximating X-ray scatter using a smooth bivariate B-spline and directly inferred spline coefficients using a lean convolutional encoder architecture \cite{roser2020deep}.
This (a) allowed us to reduce the number of parameters and computational complexity tremendously, (b) ensured that no high-frequency details of diagnostic value can get manipulated, while (c) still meeting the high accuracy of the U-net.

\subsection{Contributions}
In this work, we extend the idea of approximating X-ray scatter as bivariate B-spline by integrating its evaluation into the computational graph of a neural network.
We analyze our approach in depth and compare it to several U-net realizations in a nested cross-validation study using synthetic head and thorax X-ray image data.
Besides benchmarking the parameter and computational complexity, we investigate all networks in terms of their frequency response, the power spectral density of the co-domain, and the response to unseen noise levels.
Finally, we demonstrate how our method performs when applied to real data in an anthropomorphic thorax phantom study.

\section{Materials and Methods}
\subsection{X-ray Projection Formation using B-splines}
In general, in both techniques, X-ray fluoroscopy and CBCT it is assumed that primary photons either reach the detector along a straight line or are absorbed completely.
Therefore, the observed X-ray projection is typically described in terms of its primary component $\vec{I}_\text{p} \in \mathbb{R}^{w \times h}$ with image width $w$ and height $h$ in pixels.
The intensity at pixel $(u, v)$ is given by the Beer-Lambert law
\begin{equation}
    {I}_\text{p} \left(u, v\right) = \int_E {I}_0\left(u, v, E \right) e^{ -\int_\lambda \mu \left( \vec{s} + \lambda \cdot \vec{r}\left(u, v\right), E \right) d\lambda} dE \enspace .
\end{equation}
The polychromatic flat-field projection or X-ray spectrum $\vec{I}_0$ as well as the linear attenuation coefficient $\mu\left(\vec{l}, E\right)$ depend on the photon energy $E$.
Note that, due to the cone-beam geometry, $\vec{I}_0$ decreases towards the borders and thus depends on the pixel position $(u, v)$.
The linear photon attenuation further depends on the media along the straight line $\vec{l}: \mathbb{R} \mapsto \mathbb{R}^3$ from the X-ray source $\vec{s} \in \mathbb{R}^3$ in direction $\vec{r} \in \mathbb{R}^3$ to the pixel $(u,v)$.
Unfortunately, in reality, besides being photoelectrically absorbed, X-ray photons also undergo Compton scattering, Rayleigh scattering, or multiple occurrences of both effects.
Therefore, a more realistic formation model $\vec{I}$ adds scattered photons $\vec{I}_\text{s}$ leading to
\begin{equation}
   {I} \left(u, v\right) = {I}_\text{p} \left(u, v\right) + {I}_\text{s} \left(u, v\right) \enspace .
\end{equation}
Since, for CBCT, we are mostly interested in the low-frequency components of the scatter, we neglect photon shot noise and detector noise below.

In a recent study \cite{roser2020deep}, we experimentally showed that the main scatter components, which are low-frequency for diagnostic X-rays, can be well approximated by sparse bivariate B-splines with error rates below \SI{1}{\percent}.
The domain and co-domain of a B-spline of degree $k$ and order $n = k + 1$ are fully characterized by the knot vectors $\vec{t}_u = (t_{u, 1},t_{u, 2}, \dots t_{u, w_c-n} )$ and $\vec{t}_v = (t_{v, 1},t_{v, 2}, \dots t_{v, h_c-n} )$, as well as the coefficient matrix $\vec{C} \in \mathbb{R}^{w_\text{c} \times h_\text{c}}$ with width $w_\text{c}$ and height $h_\text{c}$, respectively.
Based on this, we can approximate the spatial scatter distribution as a tensor product B-spline $\tilde{\vec{I}}_{\text{s},n}$
\begin{equation}
    \tilde{I}_{\text{s},n}\left(u, v\right) = \sum_{i=1}^{w_\text{c}} \sum_{j=1}^{h_\text{c}} C\left(i,j\right) B_{i,n,\vec{t}_{u}}\left(u\right) B_{j,n,\vec{t}_{v}}\left(v\right) \,,
\end{equation}
with the basis splines $B$, which are zero for knots that do not affect the spline at the pixel $(u, v)$.
The basis splines are recursively defined by
\begin{equation}
\begin{aligned}
    B_{i,n,\vec{t}}(x) 
    & = {} & \frac{x - t_{i}}{t_{i+n} - t_{i}} & \cdot B_{i,n-1,\vec{t}}(x) \\ 
    & + {} & \frac{t_{i+n+1} - x}{t_{i+n+1} - t_{i+1}} & \cdot B_{i+1,n-1,\vec{t}}(x) \enspace,
\end{aligned}
\end{equation}
with
\begin{equation}
    B_{i,0,\vec{t}}(x) =
    \begin{cases}
        1, & \text{if } t_i \leq x < t_{i+1} \\
        0, & \text{otherwise}
    \end{cases} \enspace.
\end{equation}

Since the basis functions $B$ are equal to zero for coefficients that do not contribute to a pixel $(u, v)$, only a $n \times n$ sub-grid of $\vec{C}$ needs to be considered for each pixel.
Thus, using matrix notation, the tensor product can be reformulated to
\begin{equation}
    \tilde{I}_{\text{s},n}\left(u, v\right) = \vec{u}^\text{T} \cdot \left[\vec{M}_{n,\vec{t}_u}\left(u\right)\right] \cdot \vec{C}_\text{uv} \cdot \left[\vec{M}_{n,\vec{t}_v}\left(v\right)\right]^\text{T} \cdot \vec{v} \enspace , 
\end{equation}
with the coefficient patch $\vec{C}_\text{uv} \in \mathbb{R}^{n \times n}$ impacting the pixel $(u,v)$, the general matrix representation of a univariate B-spline $\vec{M}_{n,\vec{t}}\left(\cdot\right) \in \mathbb{R}^{n \times n}$ \cite{qin2000general}, and the vectors $\vec{u} = \begin{pmatrix} u^0 & u^1 & \cdots & u^k \end{pmatrix}^\text{T}$ and $\vec{v} = \begin{pmatrix} v^0 & v^1 & \cdots & v^k \end{pmatrix}^\text{T}$.
The general matrix representation of B-splines allows for arbitrarily spaced knot grids, i.e., in our case, it allows for endpoint interpolation, which exposes a more convenient behavior towards the borders of the evaluation grid.
Besides the borders of the grid, we currently limit our approach to uniform knot grids and cubic B-splines ($k=3$).
Thus, for most pixels $(u,v)$, the B-spline is evaluated using
\begin{equation}
    \vec{M}_{4} = \frac{1}{6} \begin{pmatrix} 1 & 4 & 1 & 0 \\ -3 & 0 & 3 & 0 \\ 3 & -6 & 3 & 0 \\ 1 & 3 & -3 & 1\end{pmatrix} \enspace .
\end{equation}

With using a fixed knot grid and evaluation grid, we can pre-calculate both $\vec{u}^\text{T} \cdot \left[\vec{M}_{n,\vec{t}_u}\left(u\right)\right]$ and $\vec{v}^\text{T} \cdot \left[\vec{M}_{n,\vec{t}_v}\left(v\right)\right]$ for each pixel $(u, v)$.
Padding the resulting vectors with zeros to account for all coefficients in $\vec{C}$ and stacking them row-wisely, we obtain the evaluation matrices $\vec{U}_n \in \mathbb{R}^{w \times w_\text{c}}$ and $\vec{V}_n \in \mathbb{R}^{h \times h_\text{c}}$, respectively.
Thus, we can calculate the cubic ($n=4$) scatter distribution given the B-spline coefficients $\vec{C}$ via
\begin{equation}
    \tilde{\vec{I}}_{\text{s},4} = \vec{U}_4 \cdot \vec{C} \cdot \vec{V}_4^\text{T} \enspace .
\end{equation}
Since neural networks can extract scatter distributions from a measured X-ray projection either directly \cite{Maier:2019:DSE} or indirectly via B-spline coefficients \cite{roser2020deep} we aim to combine deep learning with the matrix evaluation scheme.
Using the Kronecker product `$\otimes$', the derivative $\frac{\partial \tilde{\vec{I}}_{\text{s},4}}{\partial \vec{C}}$ is
\begin{equation}
    \frac{\partial \tilde{\vec{I}}_{\text{s},4}}{\partial \vec{C}} = \vec{V}_4 \otimes \vec{U}_4 \enspace .
\end{equation}
Thus, we can straightforwardly embed the B-spline evaluation into a computational graph $f_{\vec{\theta}}: \mathbb{R}^{w \times h} \mapsto \mathbb{R}^{w \times h}$ with parameters to train $\vec{\theta}$ without breaking differentiability and thus back-propagation.

\subsection{Network Architecture}
\label{sec:methods:arch}
\begin{figure}
    \centering
    \includegraphics[trim=3.5cm 0 2cm 0,width=5.5cm]{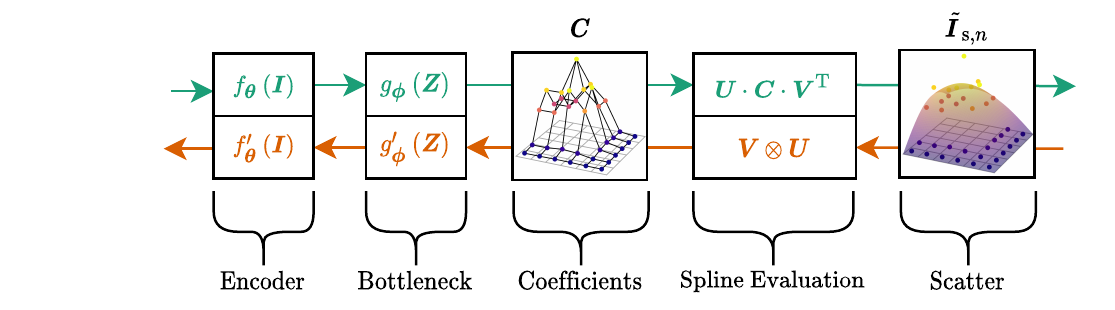}
    \caption{
        The general architecture of our proposed approach. 
        First, a (convolutional) encoder extracts the latent variables $\vec{Z}$ from the input X-ray projection $\vec{I}$.
        Second, a bottleneck network maps the latent space to bivariate spline coefficients $\vec{C}$.
        The different choices for both networks are discussed later in more detail.
        Third, we obtain the scatter estimate $\tilde{\vec{I}}_{\text{s},n}$ by evaluating the spline coefficients $\vec{C}$ using the pre-calculated evaluation grid defined by the pre-computed matrices $\vec{U}$ and $\vec{V}$.
        The top (green) path refers to the forward pass, and the bottom (orange) path to the backward pass, respectively.}
    \label{fig:general-architecture}
\end{figure}
In the following, we describe the devised neural network.
The overall architecture comprises a generic convolutional encoder followed by a bottleneck network to infer spline coefficients from measured X-ray projections, as proposed in our previous study \cite{roser2020deep}.
This architecture is depicted in Fig. \ref{fig:general-architecture}.
We employ a lean convolutional encoder $f_{\vec{\theta}}: \mathbb{R}^{w \times h} \mapsto \mathbb{R}^{w_\text{c} \times h_\text{c}}$ that consists of $d$ convolutional blocks.
A block comprises two convolutional layers with $c$ feature channels and $3 \times 3$ kernels.
Each convolutional layer is followed by a rectified linear unit (ReLU) \cite{glorot2011relu} activation, and between two blocks, $2\times 2$ average pooling is applied.
As X-ray scatter is low-frequency and to potentially limit the computational complexity of our model, we allow for additional pooling layers before the first convolutional block.
The encoder $f_{\vec{\theta}}$ is completed by a $1 \times 1$ convolution to build the weighted sum of all channels $c$.
Overall, based on convolutions only, $f_{\vec{\theta}}$ only encodes a local scatter representation $\vec{Z}$, since its receptive field does not necessarily cover the whole input image.
To establish a global context, we employ different types of bottleneck networks $g_{\vec{\phi}}$ to map $\vec{Z}$ to spline coefficients $\vec{C}$: (1) a constrained weighting matrix $\vec{W}$ with $W_{i,j} > 0$, (2) an unconstrained fully-connected layer with ReLU activation (which merely relates to an unconstrained weighting matrix with enforced non-negativity constraint), (3) two fully-connected layers with ReLU activation, or (4) two additional convolutional blocks followed by a fully-connected layer.

\subsection{Synthetic Dataset}
Acquiring raw scatter-free X-ray projections and their scatter-contaminated counterparts is tedious and time-consuming, especially at the scale to carry out deep learning.
As a solution, we leveraged the X-ray transport code MC-GPU \cite{Badal:09:MCGPU} to generate artificial pairs of scatter-contaminated and scatter-free X-ray projections.
We used openly available CT scans from The Cancer Imaging Archive (TCIA) \cite{Clark:13:TCIA} as inputs to the simulation.
In order to keep the simulation time manageable, we selected 20 head scans from the HNSCC-3DCT-RT dataset \cite{Bejanero:18:Heads} and 15 thorax scans from the CT Lymph Nodes dataset \cite{Roth:15:Lymphs}, respectively.
To prepare the phantoms for MC simulation, we employed a basic pre-processing pipeline \cite{roser2020fully} based on tissue and density estimation \cite{Schneider:00:HU2Tissue}, and connected component labeling \cite{He:17:CCL}. 
For each CT scan, we simulated a stack of 260 X-ray projections ($w=1152$, $h=768$) over an angular range of  \SI{200}{\degree}.
The source-to-isocenter and source-to-detector distances are \SI{785}{\mm} and \SI{1300}{\mm}, respectively.
Per X-ray projection, we simulated \num{5e10} photons sampled from an \SI{85}{\kilo\volt} peak voltage tungsten spectrum.
All projections were flat-field normalized.
Note that we used the data as is, without registering similar anatomies in a common reference frame.
To be more comparable to the related work \cite{Maier:2019:DSE}, speed-up training times, and suppress simulation noise, we down-sampled the projections to $384 \times 256$ pixels.
Before the down-sampling, we applied Gaussian filtering to the primary and scatter projections independently ($\sigma_\text{p}=2$, $\sigma_\text{s}=30$).
Corresponding cross-sectional slices were reconstructed on an isotropic \num[exponent-base=256]{e3} grid with \SI{1}{\mm\cubed} voxels.

\subsection{Real Dataset}
To evaluate the proposed method on real data, we scanned the thorax of an anthropomorphic phantom (PBU-60, Kyoto Kagaku Co.,\,Ltd., Kyoto, Japan) using a C-arm CBCT system (ARTIS icono floor, Siemens Healthineers AG, Forchheim, Germany).
In total, we acquired 12 datasets, three short scans for each full view grid (referred to as full), and maximum collimation (referred to as slit), both with and without an anti-scatter grid.
Each short scan consists of 397 projections (\num{648 x 472} pixels, \SI{616}{\micro\meter} isotropic spacing) over an angular range of \SI{197.5}{\degree} using \SI{85}{\kilo\volt} peak tube voltage.
The source-to-isocenter and source-to-detector distances are \SI{750}{\mm} and \SI{1200}{\mm}, respectively.
We reconstructed \num{512 x 512} slices on an \SI{484}{\micro\meter} isotropic voxel grid using an in-house reconstruction pipeline.
We regard the slit scan in conjunction with the anti-scatter grid as ground truth.

\section{Experiments}
In the following, we describe the experiments carried out to evaluate our proposed method.
Since the U-net based approach outperformed other computational scatter estimation methods by far \cite{Maier:2019:DSE}, we considered different configurations of the U-net as a baseline method.
To account for our relatively small training corpora, we evaluated our method and the baseline using a nested cross-validation approach.
For the head dataset (20 subjects), we used a $4^*3$-fold cross-validation approach and, for the thorax dataset (15 subjects), we used a $5^*4$-fold cross-validation approach.
The real dataset was only used for testing.
To keep the training procedure manageable, we divided the evaluation into meta-parameter search based on the scatter mean absolute percentage error (MAPE), and further in-depth analysis.

\subsection{Meta-Parameter Search}
Since clinic CBCT systems usually have fine-tuned acquisition protocols for each anatomic region, we evaluated each network architecture for head and thorax data separately.
To fix the meta-parameters, we only used the synthetic head dataset and the scatter MAPE as metric. 
We validated both our approach and the U-net for different combinations of depths $d$ and feature channels $c$ using Glorot initialization \cite{glorot2010understanding}, which was also used by the baseline \cite{Maier:2019:DSE}.
Furthermore, we distinguished between a deep U-net (DU-net) and a shallow U-net (SU-net), in which the number of feature maps is not doubled at each level.
For all experiments in this section, we performed a $4^*3$-fold cross-validation and trained all networks using the adaptive moments optimizer (Adam) \cite{kingma2014adam} with an initial learning rate of \num{e-4} for 100 epochs.
In total, we trained 12 networks for each configuration and used the averaged MAPE to assess their quality.
To reduce the total computation time, we stopped each training procedure when no significant performance increase was observed for 20 epochs.

In a first step, we scanned different parametrizations of the network architectures to find the most promising ones for in-depth comparison.
Overall, we tested both U-nets with $c = 16$ and $d \in \{ 4, 5, 6, 7 \}$, and our approach with $c = 16$, $d \in \{ 4, 5, 6 \}$, and additional pre-pooling $p \in \{0, 1, 2 \}$.
In addition, to decouple the architecture of our convolutional encoder $f_{\vec{\theta}}$ and the spline coefficient dimensionality, we investigated the four bottleneck architectures $g_{\vec{\phi}}$ as described in Sec. \ref{sec:methods:arch}.

\subsection{Qualitative and Quantitative Results}
Based on the findings of the previous experiments, we adapted the learning rate to \num{e-5} but kept the overall training routine.
We separately performed 4$^*$3-fold and 5$^*$4-fold cross-validations for the head and thorax datasets, respectively.
In addition to the scatter MAPE, we included the structural similary index (SSIM) of scatter-compensated reconstructions with respect to the ground truth in our evaluation.

\subsection{In-Depth Analysis}
\subsubsection{Spectral Analysis}
For clinical applications, data integrity is of utmost importance to ensure that automated systems do not alter diagnostically relevant content.
While the predictions of neural networks may appear reasonable at first glance, unrealistic perturbations can be unveiled by investigating the spectral properties of (a) the predicted images \cite{durall2020watch} or (b) the neural network itself.
Therefore, we first investigate all networks' performance concerning the predicted scatter distributions' power spectral density.
Second, from scatter estimation theory, we know that the scatter distribution can be recovered from the measured signal by the convolution with a so-called scatter kernel \cite{Ohnesorge:1999:KSE}.
This allows us to interpret a neural network for a specific pair of scatter distribution and X-ray signal as a filtering operation and assess its frequency response.

\subsubsection{Noise Analysis}
Assessing a network's robustness is inherently difficult given small training corpora.
Therefore, adversarial attacks are often used to expose weaknesses deliberately.
Since we trained all networks on noise-free data, testing them on data with different unseen noise levels appeared appropriate.
To this end, we applied Poisson noise associated with different photon counts ranging from \numrange{e3}{e5} to our head dataset and investigated the accuracy of the networks' predictions.

\subsubsection{Runtime Analysis}
For interventional applications, the fast execution speed of computer programs is essential.
Therefore, we benchmarked the average execution time for all networks for different batch sizes using a 12-core CPU (Intel(R) Xeon(R) Silver 4116 CPU 2.10GHz).

\subsubsection{Real Data Analysis}
To confirm the generalizability to real data, we tested the networks, which performed best on our synthetic thorax dataset, on the real thorax dataset.
For reference, we considered various different scatter suppression techniques, namely using an anti-scatter grid or slit scanning.
As an almost scatter free baseline, we considered the configuration to use both, an anti-scatter grid and slit scanning.
Note that we neither performed an intensity or a geometry calibration in between the scans to account for the missing anti-scatter grid.

\section{Results}

\subsection{Meta-Parameter Search}

\subsubsection{Network Parametrization}
\begin{figure}
    \centering
    \input{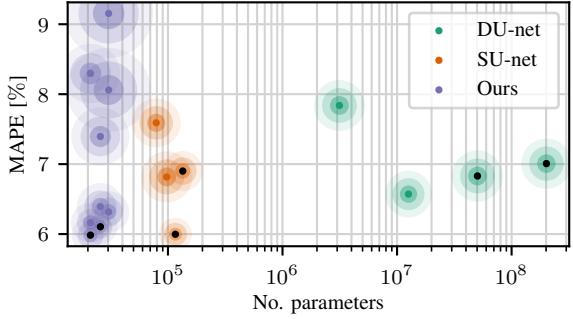}
    \caption{Distribution of different network configurations for our approach and the U-net in terms of absolute percentage errors averaged over all folds and patients. Note that, due to visualization purposes, the standard deviation encoded by the circular margins does not correspond to absolute values but rather indicates the relative spread between the networks. The standard deviation is in the range of \SIrange{1}{5}{\percent}. The x-axis refers to the number of parameters in the convolutional layers. }
    \label{fig:results:overview}
\end{figure}
Figure \ref{fig:results:overview} establishes the relationship between the number of convolutional parameters to train for each network to the averaged error rates of all folds and patients.
All networks' error rates range between \SIrange{6}{9}{\percent}, and we observed that overall the more compact networks outperform the DU-nets.
Based on these findings, we selected two spline networks ($c=16$, $(d, p) \in \{(4, 1), (5, 0)\}$) and four U-nets (deep and shallow, $c=16$, $d \in \{6, 7\}$) for further investigations.


\subsubsection{Bottleneck}
\begin{figure}
    \centering
    \input{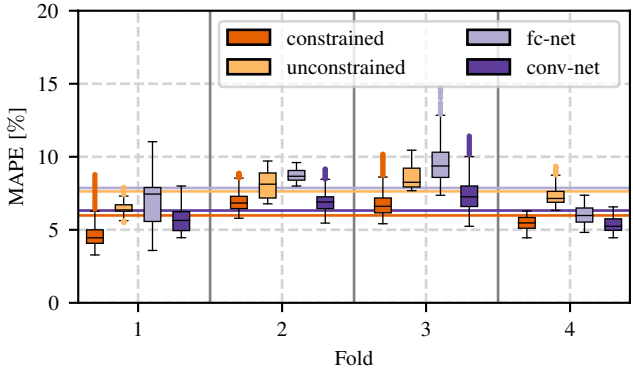}
    \caption{Boxplots of the mean absolute percentage error (MAPE) of our proposed method for different bottleneck architectures averaged over the validation folds for each training fold. The horizontal lines indicate the mean value over all test and validation folds. }
    \label{fig:results:bottleneck}
\end{figure}
\begin{figure}
    \centering
    \adjustbox{trim=0.2cm 0.6cm 0.2cm 0,clip}{\input{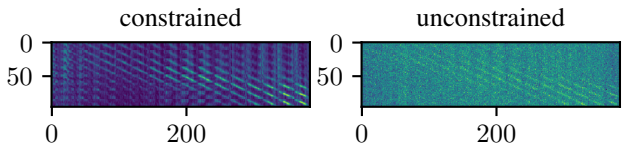}}
    \caption{Normalized constrained (left) and unconstrained (right) weighting matrices used to map the latent variables $\vec{Z}$ to spline coefficients $\vec{C}$. }
    \label{fig:results:weighting}
\end{figure}
Figure \ref{fig:results:bottleneck} shows the results for the different bottleneck architectures.
Our proposed constrained weighting matrix, homogeneously initialized, achieves the best results on average, even surpassing the convolutional bottleneck followed by a fully-connected layer.
The unconstrained fully-connected layer architectures yield considerably worse results.
Figure \ref{fig:results:weighting}, which shows the weighting matrix for the constrained and unconstrained case, substantiates this finding.
While the constrained matrix converges to a block circulant matrix, which merely relates to a convolution, the unconstrained one hardly resembles a sensible operation and is overall noisy.


\subsection{Qualitative and Quantitative Results}

\begin{figure*}[p]
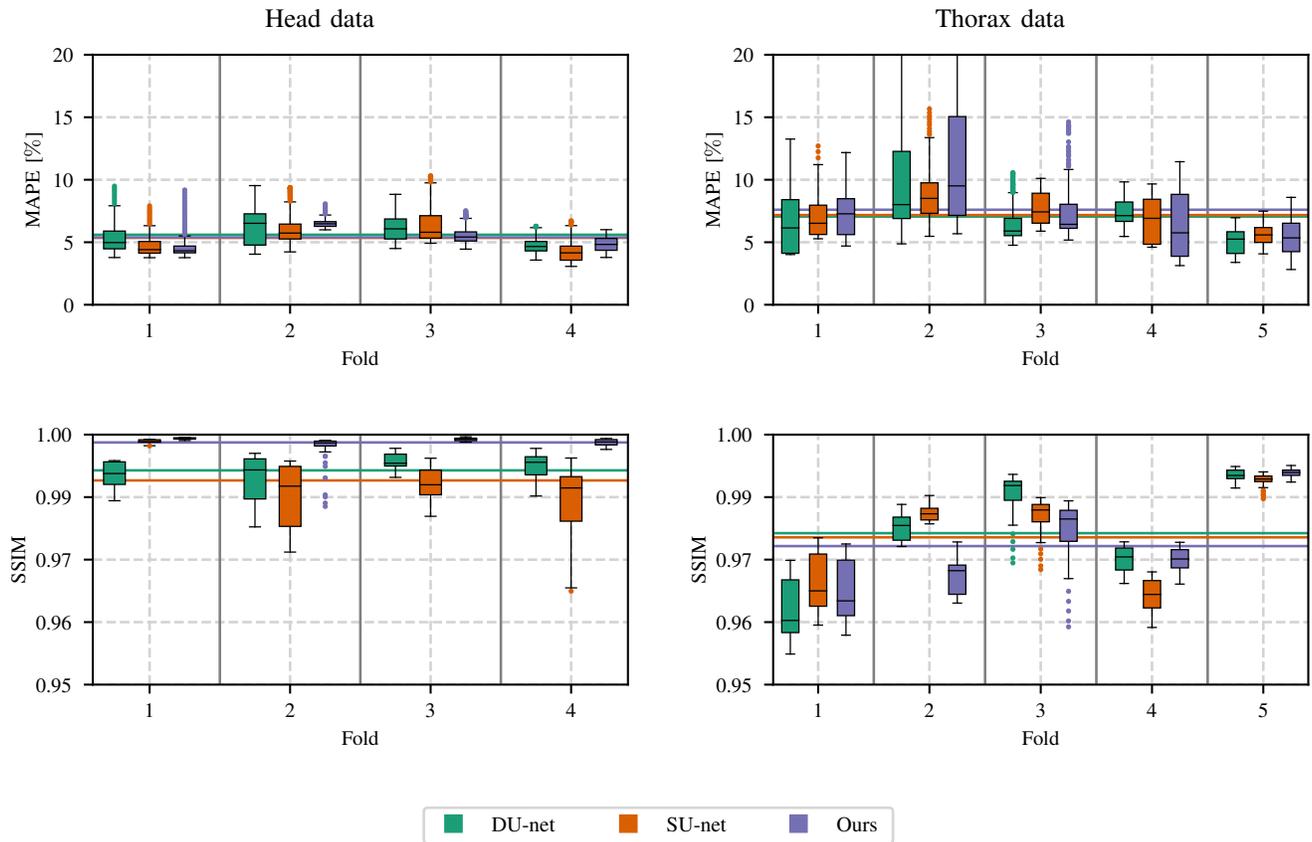

    \centering
    \begin{tabular}{c c}
    Head data & Thorax data \\ 
    \scalebox{0.97}{\input{head_mape_fold.pgf}} &
    \scalebox{0.97}{\input{thorax_mape_fold.pgf}} \\
    \scalebox{0.97}{\input{head_ssim_fold.pgf}} &
    \scalebox{0.97}{\input{thorax_ssim_fold.pgf}} \\
    \multicolumn{2}{c}{
\begingroup%
\makeatletter%
\begin{pgfpicture}%
\pgfpathrectangle{\pgfpointorigin}{\pgfqpoint{2.550000in}{0.300000in}}%
\pgfusepath{use as bounding box, clip}%
\begin{pgfscope}%
\pgfsetbuttcap%
\pgfsetmiterjoin%
\definecolor{currentfill}{rgb}{1.000000,1.000000,1.000000}%
\pgfsetfillcolor{currentfill}%
\pgfsetlinewidth{0.000000pt}%
\definecolor{currentstroke}{rgb}{1.000000,1.000000,1.000000}%
\pgfsetstrokecolor{currentstroke}%
\pgfsetdash{}{0pt}%
\pgfpathmoveto{\pgfqpoint{0.000000in}{0.000000in}}%
\pgfpathlineto{\pgfqpoint{2.550000in}{0.000000in}}%
\pgfpathlineto{\pgfqpoint{2.550000in}{0.300000in}}%
\pgfpathlineto{\pgfqpoint{0.000000in}{0.300000in}}%
\pgfpathclose%
\pgfusepath{fill}%
\end{pgfscope}%
\begin{pgfscope}%
\pgfsetbuttcap%
\pgfsetmiterjoin%
\definecolor{currentfill}{rgb}{1.000000,1.000000,1.000000}%
\pgfsetfillcolor{currentfill}%
\pgfsetfillopacity{0.800000}%
\pgfsetlinewidth{1.003750pt}%
\definecolor{currentstroke}{rgb}{0.800000,0.800000,0.800000}%
\pgfsetstrokecolor{currentstroke}%
\pgfsetstrokeopacity{0.800000}%
\pgfsetdash{}{0pt}%
\pgfpathmoveto{\pgfqpoint{0.060808in}{0.050173in}}%
\pgfpathlineto{\pgfqpoint{2.472222in}{0.050173in}}%
\pgfpathquadraticcurveto{\pgfqpoint{2.494444in}{0.050173in}}{\pgfqpoint{2.494444in}{0.072395in}}%
\pgfpathlineto{\pgfqpoint{2.494444in}{0.216222in}}%
\pgfpathquadraticcurveto{\pgfqpoint{2.494444in}{0.238444in}}{\pgfqpoint{2.472222in}{0.238444in}}%
\pgfpathlineto{\pgfqpoint{0.060808in}{0.238444in}}%
\pgfpathquadraticcurveto{\pgfqpoint{0.038585in}{0.238444in}}{\pgfqpoint{0.038585in}{0.216222in}}%
\pgfpathlineto{\pgfqpoint{0.038585in}{0.072395in}}%
\pgfpathquadraticcurveto{\pgfqpoint{0.038585in}{0.050173in}}{\pgfqpoint{0.060808in}{0.050173in}}%
\pgfpathclose%
\pgfusepath{stroke,fill}%
\end{pgfscope}%
\begin{pgfscope}%
\pgfsetbuttcap%
\pgfsetmiterjoin%
\definecolor{currentfill}{rgb}{0.105882,0.619608,0.466667}%
\pgfsetfillcolor{currentfill}%
\pgfsetlinewidth{1.003750pt}%
\definecolor{currentstroke}{rgb}{0.105882,0.619608,0.466667}%
\pgfsetstrokecolor{currentstroke}%
\pgfsetdash{}{0pt}%
\pgfsys@defobject{currentmarker}{\pgfqpoint{-0.041667in}{-0.041667in}}{\pgfqpoint{0.041667in}{0.041667in}}{%
\pgfpathmoveto{\pgfqpoint{-0.041667in}{-0.041667in}}%
\pgfpathlineto{\pgfqpoint{0.041667in}{-0.041667in}}%
\pgfpathlineto{\pgfqpoint{0.041667in}{0.041667in}}%
\pgfpathlineto{\pgfqpoint{-0.041667in}{0.041667in}}%
\pgfpathclose%
\pgfusepath{stroke,fill}%
}%
\begin{pgfscope}%
\pgfsys@transformshift{0.194141in}{0.155111in}%
\pgfsys@useobject{currentmarker}{}%
\end{pgfscope}%
\end{pgfscope}%
\begin{pgfscope}%
\definecolor{textcolor}{rgb}{0.000000,0.000000,0.000000}%
\pgfsetstrokecolor{textcolor}%
\pgfsetfillcolor{textcolor}%
\pgftext[x=0.394141in,y=0.116222in,left,base]{\color{textcolor}\rmfamily\fontsize{8.000000}{9.600000}\selectfont DU-net}%
\end{pgfscope}%
\begin{pgfscope}%
\pgfsetbuttcap%
\pgfsetmiterjoin%
\definecolor{currentfill}{rgb}{0.850980,0.372549,0.007843}%
\pgfsetfillcolor{currentfill}%
\pgfsetlinewidth{1.003750pt}%
\definecolor{currentstroke}{rgb}{0.850980,0.372549,0.007843}%
\pgfsetstrokecolor{currentstroke}%
\pgfsetdash{}{0pt}%
\pgfsys@defobject{currentmarker}{\pgfqpoint{-0.041667in}{-0.041667in}}{\pgfqpoint{0.041667in}{0.041667in}}{%
\pgfpathmoveto{\pgfqpoint{-0.041667in}{-0.041667in}}%
\pgfpathlineto{\pgfqpoint{0.041667in}{-0.041667in}}%
\pgfpathlineto{\pgfqpoint{0.041667in}{0.041667in}}%
\pgfpathlineto{\pgfqpoint{-0.041667in}{0.041667in}}%
\pgfpathclose%
\pgfusepath{stroke,fill}%
}%
\begin{pgfscope}%
\pgfsys@transformshift{1.109347in}{0.155111in}%
\pgfsys@useobject{currentmarker}{}%
\end{pgfscope}%
\end{pgfscope}%
\begin{pgfscope}%
\definecolor{textcolor}{rgb}{0.000000,0.000000,0.000000}%
\pgfsetstrokecolor{textcolor}%
\pgfsetfillcolor{textcolor}%
\pgftext[x=1.309347in,y=0.116222in,left,base]{\color{textcolor}\rmfamily\fontsize{8.000000}{9.600000}\selectfont SU-net}%
\end{pgfscope}%
\begin{pgfscope}%
\pgfsetbuttcap%
\pgfsetmiterjoin%
\definecolor{currentfill}{rgb}{0.458824,0.439216,0.701961}%
\pgfsetfillcolor{currentfill}%
\pgfsetlinewidth{1.003750pt}%
\definecolor{currentstroke}{rgb}{0.458824,0.439216,0.701961}%
\pgfsetstrokecolor{currentstroke}%
\pgfsetdash{}{0pt}%
\pgfsys@defobject{currentmarker}{\pgfqpoint{-0.041667in}{-0.041667in}}{\pgfqpoint{0.041667in}{0.041667in}}{%
\pgfpathmoveto{\pgfqpoint{-0.041667in}{-0.041667in}}%
\pgfpathlineto{\pgfqpoint{0.041667in}{-0.041667in}}%
\pgfpathlineto{\pgfqpoint{0.041667in}{0.041667in}}%
\pgfpathlineto{\pgfqpoint{-0.041667in}{0.041667in}}%
\pgfpathclose%
\pgfusepath{stroke,fill}%
}%
\begin{pgfscope}%
\pgfsys@transformshift{2.000016in}{0.155111in}%
\pgfsys@useobject{currentmarker}{}%
\end{pgfscope}%
\end{pgfscope}%
\begin{pgfscope}%
\definecolor{textcolor}{rgb}{0.000000,0.000000,0.000000}%
\pgfsetstrokecolor{textcolor}%
\pgfsetfillcolor{textcolor}%
\pgftext[x=2.200016in,y=0.116222in,left,base]{\color{textcolor}\rmfamily\fontsize{8.000000}{9.600000}\selectfont Ours}%
\end{pgfscope}%
\end{pgfpicture}%
\makeatother%
\endgroup%
} \\
    \end{tabular}
    \caption{Quantitative boxplots for our synthetic datasets. The top row shows the mean absolute percentage errors (MAPE) with respect to the scatter ground truth for each test fold. The bottom row shows structural similarity indices (SSIM) between reconstructed volumes from the simulated ideal primary signal and the scatter-corrected ones using neural networks. The horizontal lines indicate the overall average across all folds. For the sake of clarity, we only depict the best performing networks in each category. For the head dataset, we depict the DU-net with $d=6$, SU-net with $d=7$, and our method with $d=4$ and $p=1$. For the thorax dataset, we depict the DU-net with $d=7$, SU-net with $d=6$, and our method with $d=5$ and $p=0$.}
    \label{fig:results:quantitative}
\end{figure*}
\begin{figure*}[p]
    \centering
    \includegraphics[trim=0 1cm 0 0,clip]{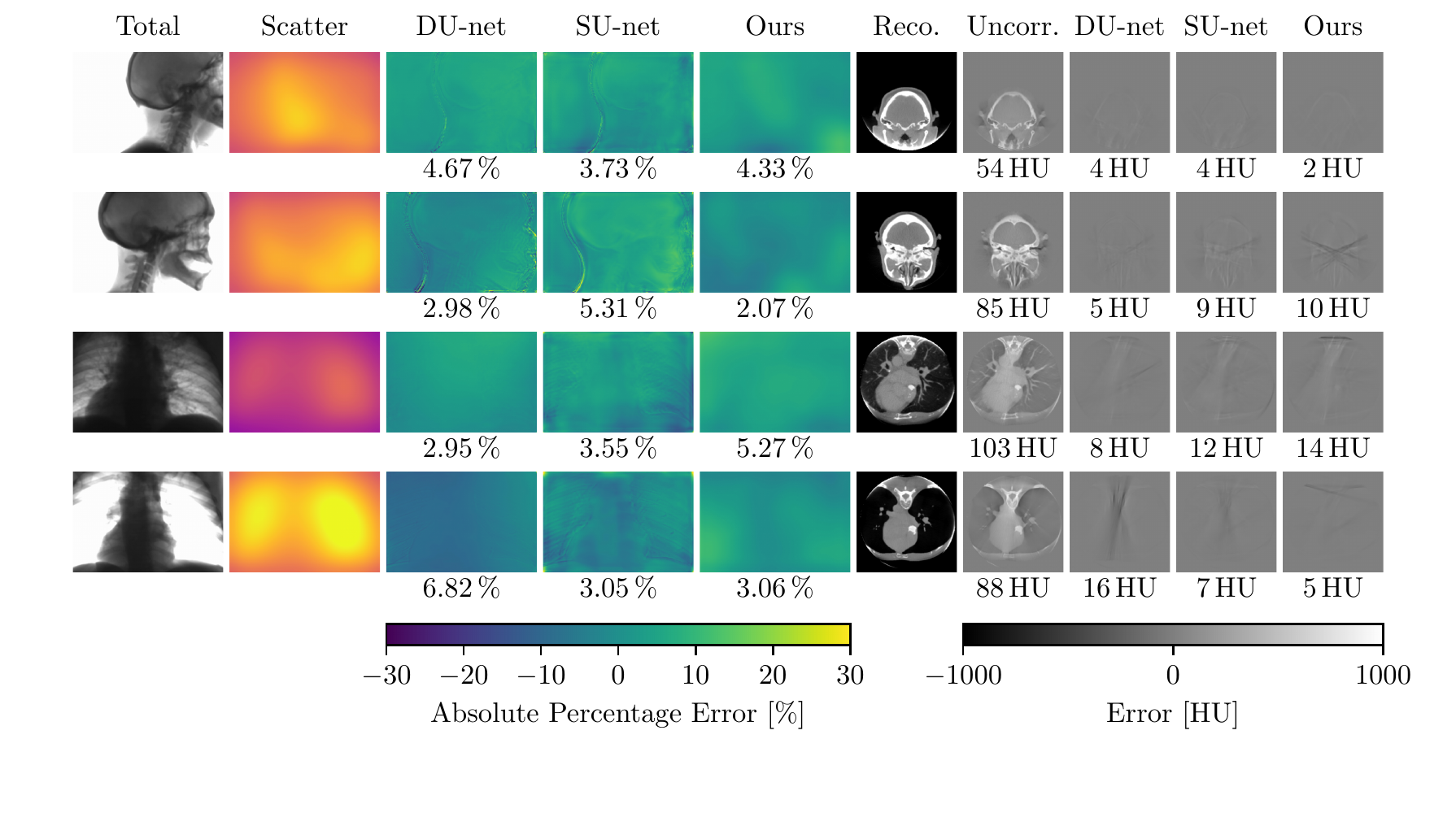}
    \caption{Selected subjects of both datasets. The results of the learning-based approaches are given in terms of error maps using the absolute percentage error (APE) for scatter and the absolute error (AE) for the reconstructed slices. For convenience, the associated mean APE (MAPE) and mean AE (MAE) are also provided below each output.}
    \label{fig:results:qualitative}
\end{figure*}
Figure \ref{fig:results:quantitative} shows fold-wise boxplots for the two synthetic datasets.
We observe similar error rates of approximately \SI{5}{\percent} across all folds and network configurations regarding the predicted scatter distributions for the head dataset.
In general, all networks achieve high SSIM values above 0.99 when comparing the reconstructed volumes from the simulated ideal primary signal to their counterparts obtained using the scatter-corrected projections.
Note that for the head data, our approach performed equally well for all folds, whereas the results obtained with the U-nets varied more widely.
Processing the thorax datasets, on the other hand, was more challenging.
Again, all networks performed comparably well with error rates of approximately \SI{7.5}{\percent}.
In comparison to the head dataset, larger error margins and outliers were found.
This can also be seen with the SSIM, which is, on average, just below 0.98.
Again, we find that our proposed method is on par with U-net-like structures from a quantitative point of view.

Looking at selected subjects of both datasets in Fig. \ref{fig:results:qualitative} reveals the potential advantages of the proposed approach.
Modeling the predicted scatter as a B-spline intrinsically limits the network output to smooth surfaces.
In contrast, both U-nets preserve some details of the input, especially the shallow architecture.
However, the reconstructed slices show no systematic trend, and all methods can adequately recover the desired signal. 

\subsection{In-Depth Analysis}
\subsubsection{Spectral Analysis}
\begin{figure}
    \centering
    \input{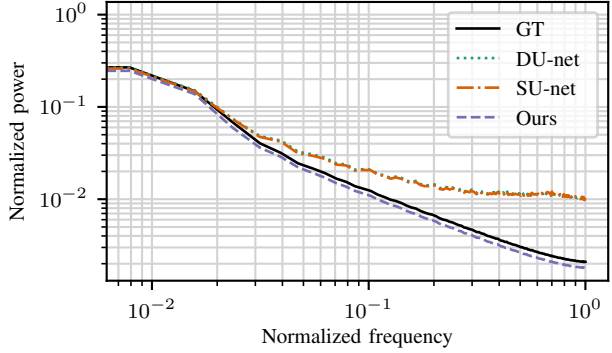}
    \caption{Power spectral density plots for the X-ray scatter distributions of the ground truth (black solid), ours (light orange dashed), and the U-net (light purple dotted) averaged over all test folds. Note that the graphs associated with both U-nets are almost identical.}
    \label{fig:results:psd}
\end{figure}
We calculated the power spectral density by azimuthally averaging the magnitudes of the 2D Fourier transform of X-ray scatter distributions.
Also, we averaged all power spectral densities of all projection, patients, and test folds.
The resulting densities plotted in Fig. \ref{fig:results:psd} support our previous findings.
While the U-nets yield numerically sound predictions, they systematically boost the high frequencies in the scatter distributions.
Our approach, however, preserves the real power spectral density over the whole frequency spectrum.
\begin{figure}
    \centering
    \input{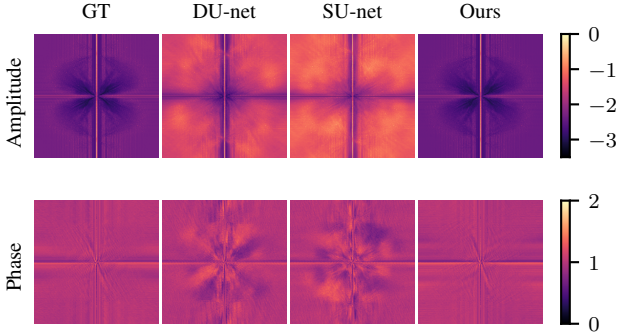}
    \caption{Frequency responses of the systems for one test patient averaged over all validation folds. From top to bottom: Normalized $\log$-amplitude, phase in radians. From left to right: ideal system (GT), our spline-net (Ours), deep U-net (DU-net), shallow U-net (SU-net).}
    \label{fig:results:response}
\end{figure}
As mentioned above, conventional convolutional neural networks can be interpreted for a single input image in terms of a filtering operation.
Thus, we divided the Fourier transform of the output by the Fourier transform of the input to obtain the respective frequency responses.
Figure \ref{fig:results:response} shows the amplitude and phase of an ideal system's frequency responses, for our method and both U-nets, averaged over all projections for one patient.
Our spline-net's frequency response was closer to the ideal frequency response in both, amplitude and phase. 
However, we observed a noticeable intensity shift in our method. 
The U-nets, in contrast, both exhibited larger deviations in the patterns of amplitude and phase, indicating that their represented operation is less predictable.

\subsubsection{Noise Analysis}
\begin{figure}
    \centering
    \input{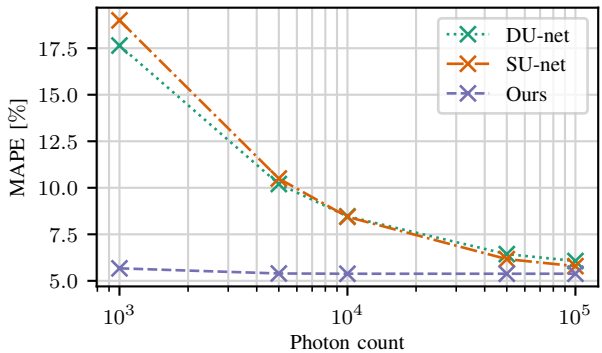}
    \caption{Error rates of predicted scatter distributions for different noise levels averaged over all validation and test fold networks. All networks have been trained with noise-free data. Note that a lower photon count relates to a higher noise level.}
    \label{fig:results:noise}
\end{figure}
Figure \ref{fig:results:noise} shows plots of the networks' error rates when confronted with noise.
We could confirm that the U-net is very sensitive to unseen noise levels in both configurations, whereas our approach performs more robustly.

\subsubsection{Runtime Analysis}
\begin{figure}
    \centering
    \input{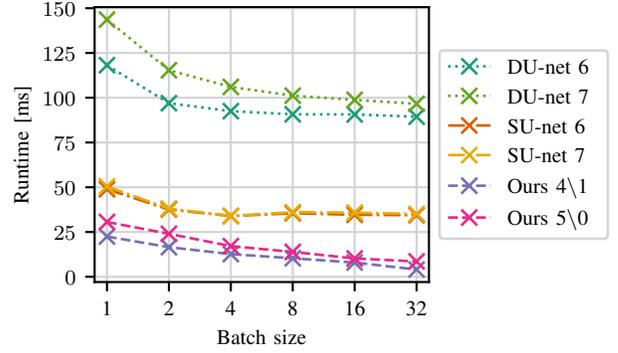}
    \caption{Runtimes per projection in \si{\milli\second} for different architectures (specified by the depth and optional pre-pooling, $d$\,\textbackslash{}\,$p$) and batch sizes. }
    \label{fig:results:runtime}
\end{figure}
Figure \ref{fig:results:runtime} compiles the inference speed of all considered networks.
As expected from the parameter complexity, our approach is the fastest with \SIrange{4}{30}{\ms} and therefore \numrange{1.7}{8.5} times faster than the SU-net with \SIrange{34}{50}{\ms}. The DU-net is the slowest with \SIrange{89}{144}{\ms}.

\subsubsection{Real Data Analysis}
\begin{figure*}
    \centering
    \input{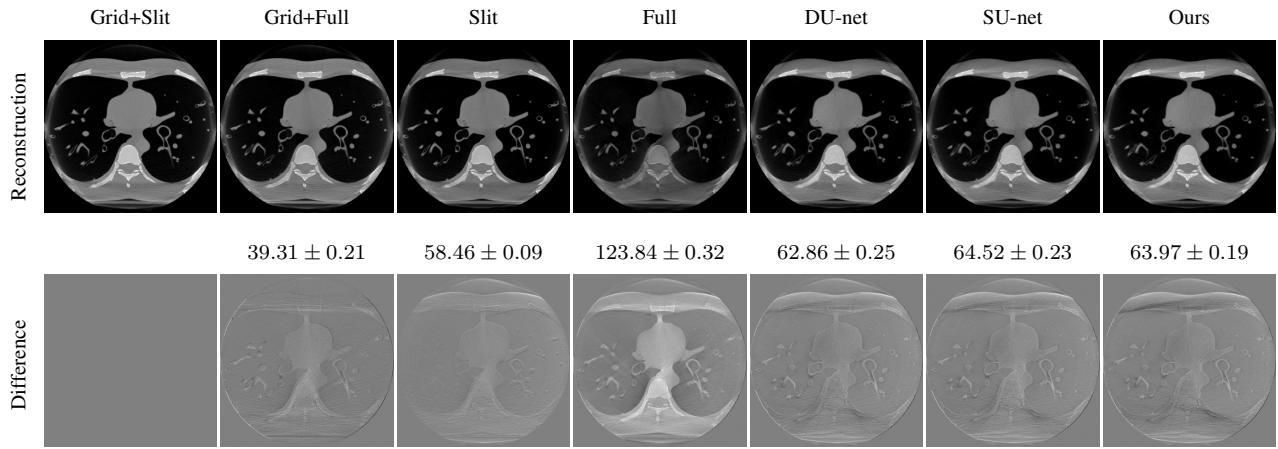}
    \caption{Central slices of reconstructed volumes for different scatter compensation strategies and errors taken with respect to the reconstruction result resulting from the grid + slit data acquisition. Grid refers to employing a conventional anti-scatter grid. Slit refers to the most narrow collimator setting available on the X-ray imaging system. Full refers to using no collimation at all. Both, the reconstructed slices as well as the difference images are shown using a gray level window $[-1000, 1000]$ HU. For convenience the absolute average HU errors and the associated standard deviations with respect to three consecutive measurements are provided. }
    \label{fig:results:real-data}
\end{figure*}
As shown in Fig. \ref{fig:results:real-data}, all methods were able to compensate for scatter artifacts well.
However, employing an anti-scatter grid still yielded the lowest overall error in Hounsfield units (HU) as compared to using slit collimation in addition.
The learning-based approaches performed about as well as the slit scanning without an anti-scatter grid, which is in accordance to previous findings \cite{Maier:2019:DSE}.
Overall, the networks achieved similar error rates, and no systematic trend was observable.

\section{Discussion}
\label{sec:discussion}
X-ray scatter is a major source of artifacts in interventional CBCT.
Deep-learning-based approaches have shown the potential to outperform conventional physical or algorithmic approaches to scatter compensation.
Without incorporating prior information, scatter distributions can be inferred from the measured X-ray signal  \cite{Maier:2019:DSE}.
However, data integrity and robustness are critical aspects of clinical imaging, which can be violated by deep neural networks \cite{huang2018some}.

To ensure sound scatter estimates, we proposed to embed bivariate B-splines in neural networks to constrain their co-domain to smooth results \cite{roser2020deep}.
By reformulating the spline evaluation in terms of matrix multiplications, we were able to integrate B-splines with neural networks without further ado such that end-to-end training was feasible.
In an extensive cross-validation using synthetic data, we showed that our proposed lean convolutional encoder using B-spline evaluation performs on par with several U-net based architectures.
We substantiated this finding in a first phantom study.
There, our approach performed basically as well as the U-net and the slit scanning technique (without anti-scatter grid), which was used as a baseline before \cite{Maier:2019:DSE}.

Note, however, that the proposed method offers several advantages not present in the U-net architectures.
First, we considerably lowered the parameter and runtime complexity, rendering our method suitable for a variety of hardware.
More importantly, we verified that our spline-based approach ensures data integrity concerning the power spectral density of scatter estimates and the overall network's frequency response.
This property ensures that no high-frequency details, which relate to anatomic structures or pathologies, are altered.
In comparison, the U-net considerably changes the concerning spectral contents, which was already shown for neural networks containing up-convolutions \cite{durall2020watch}.
As our spline network corresponds to a low-pass filtering operation, it is robust towards noise even when trained on noise-free data.

While we implemented the U-net baselines to the best of our knowledge, our error rates are higher than previously reported \cite{Maier:2019:DSE}.
Potential reasons include but are not limited to (a) different simulation codes, (b) our smaller training corpora, and (c) the heterogeneity of our data.
Our simulation setup currently assumes an ideal detector, and we do not consider the domain shift between synthetic and real data.

For future work, we indicate several research directions for either method or data and experiments.
First, since our preferred bottleneck weighting matrix converges to a block circulant matrix, a fixed representation is desirable to further reduce the number of trainable parameters and increase the plausibility of our approach.
For instance, training both the encoder and the bottleneck separately or introducing additional constraints are promising approaches to do so.
Second, replacing the bottleneck fully-connected layer with a small U-net reduces the number of parameters while still covering the entire latent space with its receptive field.
Third, since our network is end-to-end trainable, it appears reasonable to include the reconstruction into the computational graph to calculate the loss function in the CBCT domain \cite{syben2019pyro,syben2020known}.
Last but not least, we believe our approach applies to low-frequency signal estimation and correction in general, e.g., bias field correction in magnetic resonance imaging, ultrasound imaging, or  microscopy techniques.

\section{Conclusion}
\label{sec:conclusion}
Embedding B-splines in neural networks ensures data integrity for low-frequency signals.
This reduces the number of network parameters needed to arrive at physically sensible results, and thanks to the reduced parameter set, network inference can be made faster.


%

\appendices

\section*{Disclaimer}
The concepts and information presented in this article are based on research and are not commercially available.



\bibliographystyle{IEEEtran}
\bibliography{references}
\end{document}